\documentclass[journal]{IEEEtran}
\usepackage{colortbl}
\usepackage[pdftex]{graphicx}
\usepackage{color, colortbl}
\definecolor{Row0Color}{rgb}{0.83, 0.83, 0.83}
\definecolor{Row1Color}{rgb}{0.95, 0.95, 0.96}
\definecolor{Row2Color}{rgb}{0.83, 0.83, 0.83}
\usepackage{balance,soul}
\usepackage{color}
\usepackage{xcolor}
\usepackage{siunitx}
\usepackage{url}
\usepackage{marginnote}
\usepackage{todonotes}

\ifCLASSINFOpdf

\else

\fi

\usepackage[cmex10]{amsmath}

\usepackage{amsmath, amsthm, amssymb, amsfonts, array}
\usepackage{graphicx}
\usepackage{ulem}
\usepackage{subcaption}

\begin{document}
\title{Benefits of Positioning-Aided Communication Technology in High-Frequency Industrial IoT}
%
%
%

\author{\IEEEauthorblockN{Elena Simona Lohan\IEEEauthorrefmark{1}, 
Mike Koivisto\IEEEauthorrefmark{1}, 
Olga Galinina\IEEEauthorrefmark{1},
Sergey Andreev\IEEEauthorrefmark{1}, 
Antti T\"olli\IEEEauthorrefmark{2},\\
Giuseppe Destino\IEEEauthorrefmark{2}, 
M\'ario Costa\IEEEauthorrefmark{3}, 
Kari Lepp\"anen\IEEEauthorrefmark{3},  
Yevgeni Koucheryavy\IEEEauthorrefmark{1}, and 
Mikko Valkama\IEEEauthorrefmark{1} 
} \\ 
\IEEEauthorblockA{\IEEEauthorrefmark{1}Tampere University of Technology, Finland},
\IEEEauthorblockA{\IEEEauthorrefmark{2}University of Oulu, Finland},
\IEEEauthorblockA{\IEEEauthorrefmark{3}Huawei Technologies Oy, Finland}
\thanks{This work was supported by the Academy of Finland (projects PRISMA and WiFiUS) and by the project TAKE-5: The 5th Evolution Take of Wireless Communication Networks, funded by Tekes. The work of the third author is supported by a personal Jorma Ollila grant from Nokia Foundation and by the Finnish Cultural Foundation.}
}

%

\markboth{IEEE Communications Magazine}
{E. S. Lohan \MakeLowercase{\textit{et al.}}: Benefits of Positioning-Aided Communication}

\maketitle

\begin{abstract}
The future of industrial applications is shaped by intelligent moving IoT devices, such as flying drones, advanced factory robots, and connected vehicles, which may operate (semi-)autonomously. In these challenging scenarios, dynamic radio connectivity at high frequencies -- augmented with timely positioning-related information -- becomes instrumental to improve communication performance and facilitate efficient computation offloading. Our work reviews the main research challenges and reveals open implementation gaps in Industrial IoT (IIoT) applications that rely on location awareness and multi-connectivity in super high and extremely high frequency bands. It further conducts a rigorous numerical investigation to confirm the potential of precise device localization in the emerging IIoT systems. We focus on positioning-aided benefits made available to multi-connectivity IIoT device operation at $28$~GHz, which notably improve data transfer rates, communication latency, and extent of control overhead.
\end{abstract}



\section{Introduction and Motivation} 

It is predicted that the Internet of Things (IoT) market will reach $561$ billion US dollars by $2022$\footnote{``Internet of Things (IoT) Market by Software Solution, Service, Platform, Application Area, and Region -- Global Forecast to 2022'', Research and Markets, June 2017. [Accessed on 08/2018]}. A major share of these revenues will come from the industrial IoT (IIoT) segment, which comprises manufacturing, transportation, logistics, and utilities. Overlapping with the paradigm of the Industry 4.0, the early definitions of the IIoT portray it as a collection of interconnected devices, sensors, and actuators that are capable of operating in (semi-)autonomous manner, are connected to the Internet (typically through low-power radio technologies), and are deployed in various industrial environments, such as factory floors, underground mines, excavation sites, oil and gas fields, and automation halls. 

More recently, the definition of the IIoT has been applied in a broader context, by encompassing a wide variety of applications in smart living, safety, surveillance, and intelligent transportation systems, including autonomous vehicles, flying drones, and industrial automated robots~\cite{Orsino2017}. Among the top use cases driving the acceleration of the IIoT, the following will play an important role: predictive maintenance of machines and robots, self-optimizing large-scale production of various goods and assets, autonomous fleets of connected cars and unmanned aerial vehicles, as well as remote patient monitoring. 

The performance criteria in the emerging IIoT applications are multi-faceted. They range from conventional traffic-centric demands for large capacity and high throughput to emerging requirements of (ultra-)low latency in mission-critical applications with stringent reliability and adequate scalability. A taxonomy underlying IIoT was recently discussed in~\cite{Schulz2017}, where a distinct separation has been made between the `massive' and the `critical' pillars of this sector. Here, the former refers to a large number of interconnected low-cost devices that require appropriate scalability, while the latter envisions that the device volumes are much smaller and the focus shifts to achieving very low latency, better robustness, and high availability of uninterrupted connectivity.

Accounting for the rapid growth pace of critical IIoT applications, an immediate goal of next-generation radio technology becomes the construction of comprehensive wireless connections among diverse stationary and mobile machines over extensive coverage areas. Today, this construction is pioneered by the fifth generation (5G) of cellular networks and the respective standardization is underway at full speed. Seamless connectivity support for unconstrained mobility is particularly challenging for mission-critical IIoT devices moving at various speeds over a particular geographical area, such as in industrial automation and control, remote manufacturing, intelligent transportation systems, and numerous other applications.

With the above background in mind, in this paper we (i) provide a concise review of new IIoT challenges related to the use of super high and extremely high frequency communication, (ii) propose prominent positioning-related technology solutions to improve the performance of mobile IIoT applications, and (iii) by relying on both ray-based modeling and system-level simulations, evaluate their performance in selected scenarios, where devices move at various speeds. The rest of this article is structured as follows. Section II summarizes the key challenges in the context of IIoT, while Section III discusses attractive and novel lines of technology development. Then, in Section IV we detail the proposed modeling methodology, and the respective numerical results follow in Section V. The work concludes in Section VI.

\section{Challenges in Emerging IIoT Systems}

\begin{figure}[!ht]
\centering
\includegraphics[width=1.0\columnwidth]{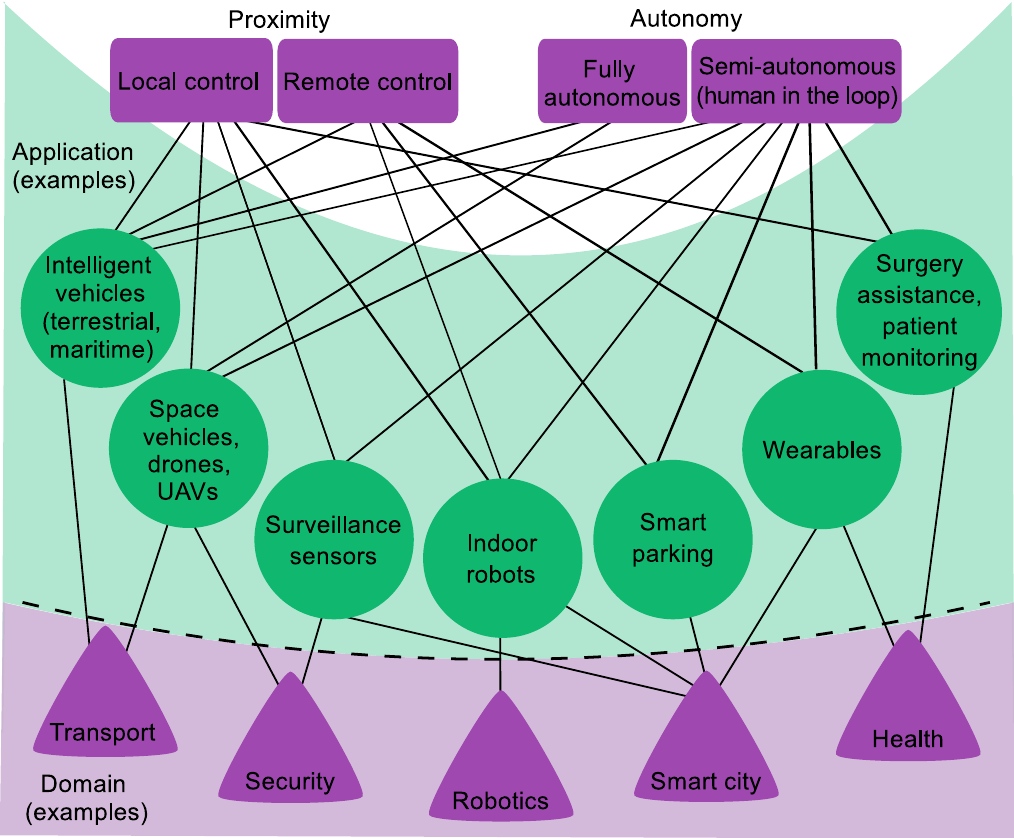}
\caption{Contemporary classification of novel IIoT applications}
\label{fig:IIoTClassif}
\end{figure}

Advanced IIoT systems can be regarded from two different perspectives, as demonstrated in Fig.~\ref{fig:IIoTClassif}: (i) their physical proximity to the central control unit or the core network node, where we differentiate between local control and remote control operations, and (ii) their level of autonomy, where one could distinguish fully-autonomous and semi-autonomous systems. While today's technology trend leans towards fully autonomous system design, the intermediate steps in this direction will include remote controlled semi-autonomous devices, such as fleets of connected vehicles or marine vessels. The envisioned application domains of the IIoT are exceptionally vast and span across multiple industries, such as transportation, healthcare, surveillance, and environmental monitoring.

As follows from Fig.~\ref{fig:IIoTClassif}, the complex and dynamic network of interconnected IIoT devices needs to be maintained and utilized efficiently. To this aim, several key requirements for critical IIoT systems emerge along the lines of (ultra-)low communication latency, seamless and reliable radio connectivity, low deployment and maintenance costs, as well as fast processing capabilities. To enjoy unconstrained mobility, future IIoT devices (cars, robots, drones, etc.), might need to remain battery-powered or battery-less (i.e., harnessing the energy of their surrounding environment or receiving it through the dedicated wireless power transfer). 

Indeed, energy dissipation of advanced IIoT equipment, such as robots and drones, becomes a growing concern as it threatens to limit the lifetime of the corresponding network infrastructures. Fortunately, recent progress in battery technology as well as in alternative power sources promises to alleviate this constraint and help achieve reliable connectivity performance. Further, the costs of providing maintenance to large numbers of advanced machines may become prohibitive, which is particularly crucial for emerging real-time control and fault diagnostics applications that involve moving industrial robots and drones.

As a result, the emerging industrial services increasingly rely on integrated indoor/outdoor deployments of moving connected machines. To facilitate their efficient operation, such systems require novel 3D positioning and location tracking mechanisms that could potentially offer service continuity and reliability on a massive scale. Operating advanced IIoT systems by relying solely on the conventional (2G--4G) broadband connectivity solutions might be insufficient, since such functioning may incur increased channel congestion and produce unpredictable response times. 

Another crucial direction is decreasing the computational burden of the individual IIoT devices, which may be achieved by offloading their more demanding tasks onto the edge/fog computing infrastructures~\cite{Sill2017}. While both edge and fog computing formulations -- unlike the cloud computing paradigm -- assume certain proximity to the target mobile device, there are inherent distinctions between the two concepts that are related to the level of separation between the nodes and their functionality~\cite{Sill2017}. 

\section{Key Technologies for Next-Generation IIoT}

One of the distinguishing features of the emerging 5G mobile systems is their reliance on the integration of the conventional (cmWave) and higher-frequency (mmWave) spectrum for improved throughput, latency, and reliability. While the vast majority of the current wireless systems employ frequencies below $6$ GHz, we expect that advanced moving devices, such as cars, drones, and industrial robots, will increasingly rely on the above-$6$ GHz bands for their operation with better spectral efficiency (i.e., $30$ bps/Hz and $15$ bps/Hz in the downlink and uplink, respectively) and reduced over-the-air latency (i.e., $<2$ ms).

Generally, going higher in frequency requires rethinking many aspects of the state-of-the-art transceiver architecture, physical layer techniques, and system control solutions to maintain seamless service continuity.

\subsection{Above-$6$ GHz transceiver technology for IIoT}
At higher carrier frequencies, the need for highly directional transmission and reception grows as well. Highly directional transmission and reception are thus becoming key to delivering substantial beamforming gains to ensure communication link reliability as well as control interference to/from other IIoT devices in dense deployments. A natural approach to implementing these important mechanisms is to utilize massive antenna arrays with beamforming capabilities at both the transmitter (TX) and the receiver (RX). As all-digital solutions are costly and power inefficient when the number of antennas becomes large, alternative low-power approaches are based on all-analog and hybrid architectures~\cite{Destino_2017}.

Analog beamforming offers a simple solution to steer and align the TX and RX beams in mmWave communication, but it is typically limited to single-stream and single-user scenarios wherein only one baseband per radio frequency chain is available at each end. A viable alternative for complex IIoT systems is to employ hybrid analog/digital precoding structures that allow multiple streams to be transmitted/received in parallel. In this configuration, appropriate weights for multiple analog arrays may be designed jointly with digital precoding and combining. However, this creates challenges related to beam alignment and beam tracking, since time/frequency resources are taken away from the actual data communication. More importantly, as far as wearables, moving robots, or drones are concerned, mobility is yet another crucial factor to consider~\cite{Ravindran2017}.

Positioning information can be employed to reduce the beam alignment overheads as well as develop proactive beamforming techniques. Although the 5G standardization process is not yet complete, it has already been envisioned that most devices will need to be localized in future 5G radio systems with at least \si{1}{m} of accuracy in at least 80\% of the cases~\cite{3GPP_TR22862}. In~\cite{Destino_2017,Shahmansoori2015}, novel results on mmWave-based positioning were demonstrated. It has been confirmed that by leveraging the availability of multiple antennas at the TX and RX as well as utilizing large signal bandwidth for periodic pilot transmission in the mmWave bands, centimeter-level positioning accuracy can be achieved. 

In cmWave bands, another important study is illustrated in~\cite{koivisto_joint_2017}, where periodic transmission of uplink pilot signals is exploited for high-accuracy positioning in a network-centric manner. Therefore, in the next generation of IIoT implementations a feasible solution is to utilize the available cmWave networks for providing the initial positioning information to improve the efficiency of mmWave communication. Reliance on existing cmWave technologies is especially attractive for designing the TX and RX beamformers for future mmWave access, thus reducing communication latency in prospective IIoT deployments.

\subsection{Location-aware multi-connectivity for improved operation}

Cooperative and coordinated multi-point transmission-reception schemes and spatial domain interference management have been studied extensively in the past, and their performance limits are well understood~\cite{Gesbert-Hanly-Huang-Shamai-Simeone-Yu-JSAC-2010}. Contemporary wireless standards already support efficient single-cell Multiple-Input Multiple-Output (MIMO) beamforming, which allows for smarter beamformer design to efficiently take advantage of multi-user diversity in the spatial domain. In connection to that, the basic operation of multi-cell MIMO transceiver processing has been covered by the 3GPP's Long Term Evolution (LTE) specifications and also includes support for fully centralized joint-processing coordinated multi-point (CoMP) transmission schemes. 

In prospective cloud radio access network (RAN) design, the joint beamforming is fully centralized, and the base stations act merely as virtual Transmission-Reception Points (TRPs). They are typically connected to a Virtual Central Unit (VCU) in the cloud over a high capacity and low latency backhaul link. The VCU may then exploit joint processing to utilize multiple TRPs for beamforming simultaneously. Most of past research efforts have not considered how multi-connectivity based resource allocation for moving IIoT devices could be augmented with precise positioning information. Importantly, such information needs to be made available in 3D space and in real-time, so that link blockage (by other IIoT machines) and self-blockage (by parts of the actual communicating machine) could be mitigated timely and without causing harmful session interruptions. 

Connectivity across multiple TRPs is assumed to be vital for robust and reliable communication, especially when real-time computation offloading onto the edge-computing infrastructure is required on the move~\cite{7925949}. An important challenge in IIoT for supporting parallel concurrent links/beams to/from multiple TRPs is the requirement for the network to track the direction (and, generally, the channel state) of each link. Unlike in legacy scenarios, where each mobile terminal is served solely by its closest TRP, dynamic multi-connectivity offers a possibility to provide parallel data streams even in the line-of-sight (LoS) conditions, given that the TRPs are separated geographically (assuming sufficient angular difference). From the MIMO processing perspective~\cite{Garcia2017}, this means that multiple beams can be utilized to increase either data-rate capacity or communication reliability.

Hybrid analog-digital beamforming architectures imposed by the use of carrier frequencies above $6$~GHz render multiuser interference management for multipoint connectivity to be particularly challenging. There is an underlying trade-off between higher data rate and better reliability, which should be understood in more detail for future IIoT systems. Another factor that impacts the choice of the reliability/rate balance point is connected with the beamwidth of each TRP link. Wider beamwidth provides lower communication rates but remains less susceptible to the loss of radio connection. On the other hand, in multi-connectivity systems, wider beams -- and hence lower rates per beam -- may be compensated by establishing multiple parallel streams between the moving IIoT device and its serving set of TRPs. 

\subsection{Positioning-aided dynamic computation offloading}

Historically, positioning and communication system architectures have been evolving disjointly in existing (cmWave) communication and navigation systems. As of today, no low-power positioning technology is able to seamlessly scale from tens of meters to few centimeters of accuracy, neither indoors nor outdoors. Advanced mmWave IIoT systems will be based on dense deployments of TRPs, narrow antenna beams, and very large bandwidths, and thus potentially have sufficient capacity to achieve very accurate positioning under acceptably low energy consumption. Building on this, industrial environments can highly benefit from accurate positioning information, since timely localization of machines, environment mapping, and mobility prediction can enable robust and proactive edge-cloud offloading of demanding computations. 

In particular, an industrial machine can push its raw multimedia data to the nearby cloud-computing infrastructure in real-time, instead of processing it locally (due to energy constraints). As a result, its human or robot operator can receive the already processed multimedia stream through the cloud infrastructure and then take the necessary action(s). This eventually results in latency-controlled and power-optimized end-to-end communication, location-aware interference mitigation, and improved throughputs delivered to highly mobile IIoT devices. It could be achieved, for example, by using geometric beamforming, also known as location-based or location-aware beamforming. In this context, security and privacy of the localization solutions become important aspects, even though they are outside the scope of this paper. A recent survey on these may be found in~\cite{Chen2017}.

As an advanced application, we envision a moving IIoT device that leverages high-bandwidth connections to offload its processing tasks to the edge-server in real-time. The novelty of this vision stems from two premises: (i) the mmWave positioning in 3D has not been addressed sufficiently so far for real-time applications and (ii) the cmWave moving object tracking and direction finding with low computational resources remain a challenging issue that is still not resolved. Future research efforts will need to concentrate on understanding the extent to which the task of a moving IIoT device (e.g., in remote control, monitoring, or diagnostics applications) can run locally vs. be offloaded onto the proximate edge-computing infrastructure over the mmWave links despite their intermittent nature (sudden blockages, higher path loss, etc.).

\section{Modeling Methodology for Moving IIoT}

\subsection{Representative urban scenario}

We begin by introducing an illustrative IIoT use case and then describe our modeling environment. A summary of selected numerical results follows in the next section with the aim to understand the benefits of positioning-aided IIoT communication at high frequencies. 

Our performance evaluation comprises two successive phases: (i) estimation of positioning quality for advanced IIoT devices, which are tracked with cmWave-based techniques that employ a positioning algorithm from~\cite{koivisto_joint_2017}, and (ii) system-level performance evaluation based on appropriate PHY abstraction and MAC representation for mmWave-based IIoT communication. By adopting this two-phase evaluation methodology, we are able to reduce the target problem of joint communication and positioning to two tractable sub-problems with lower complexity. It allows us to combine the accuracy of ray-tracing link-level modeling with the flexibility of system-level simulation.

\subsection{Positioning error estimation}

In what follows, we consider three distinct types of moving objects: \textit{drones} that are characterized by their average speed of \SI{20}{kmph} and antenna altitude of \SI{5}{m}; \textit{vehicles} with \SI{40}{kmph} mean speed and \SI{1.5}{m} antenna height; and \textit{pedestrians} having the speed of \SI{6}{kmph} on average and carrying the IIoT devices with \SI{1.2}{m}-high antennas. These target IIoT objects move along random trajectories in an urban environment captured by a typical outdoor Madrid grid as proposed by METIS\footnote{2D Madrid grid plan by METIS (see \textit{METIS-II}$\_D7.2\_V1.0.pdf$, p. 19): https://bscw.5g-ppp.eu/pub/bscw.cgi/d139814/ [Accessed on 08/2018]}. In our reference setup, wireless infrastructure is assumed to be dense (by e.g., deploying RAN nodes on top of the street furniture), thus resulting in relatively short inter-site distances (ISDs) and predominantly LoS operation.

In the first phase of our conducted evaluation, radio propagation channels are modeled with a comprehensive ray-tracing tool, which mimics all of the relevant multipath components and emulates wireless interference in a realistic manner. In our urban setup described above, the target IIoT device periodically (once in \SI{10}{ms}) transmits omnidirectional pilot signals to its neighboring TRPs. For that matter, it utilizes OFDM waveforms with 40 pilot sub-carriers and \SI{15}{kHz} of sub-carrier spacing, which span a relatively small effective bandwidth of \SI{3}{MHz}. Based on thus acquired channel state information, the TRPs receiving the signal in LoS conditions estimate its direction and time of arrival (DoA and ToA, respectively), as well as communicate this knowledge to the central processing unit. The latter relies on such data received from the two closest TRPs in LoS and calculates the resulting location estimate.

We note that two different estimation and tracking approaches are employed in the positioning phase for the purposes of their further comparison. First, in the more classical extended Kalman filter (EKF)-based positioning solution (referred to as the DoA-only EKF), only the available DoA measurements from the two closest TRPs are fused into the IIoT device's location estimates, whereas both DoA and ToA measurements are exploited by the second EKF-based positioning method (termed here the DoA\&ToA EKF). In the second positioning solution, the tagged industrial device is assumed to have a time-varying clock offset, while the TRPs are assumed to be mutually asynchronous, with the constant clock offsets~\cite{koivisto_joint_2017}.

\subsection{System-level performance evaluation}

In the second phase of our numerical assessment, we wrap the modeled trajectory of the target IIoT device into an abstraction of the cellular mmWave system by employing the appropriate MAC-layer and antenna beamforming considerations. In particular, we explicitly model the beam training phase, where the beam sweeping procedure is performed once per a dedicated time interval (e.g., \SI{1}{s} and \SI{5}{s} for the sake of comparison). For both the mmWave TRP and the IIoT device, we assume a set of uniform planar arrays -- of $8 \times 8$ and $4 \times 4$ antenna elements each -- that steer the beam in $16$ and $8$ directions, respectively~\cite{giordani2016multi}. 

\begin{figure}[!ht]
\centering
\includegraphics[width=1.0\columnwidth]{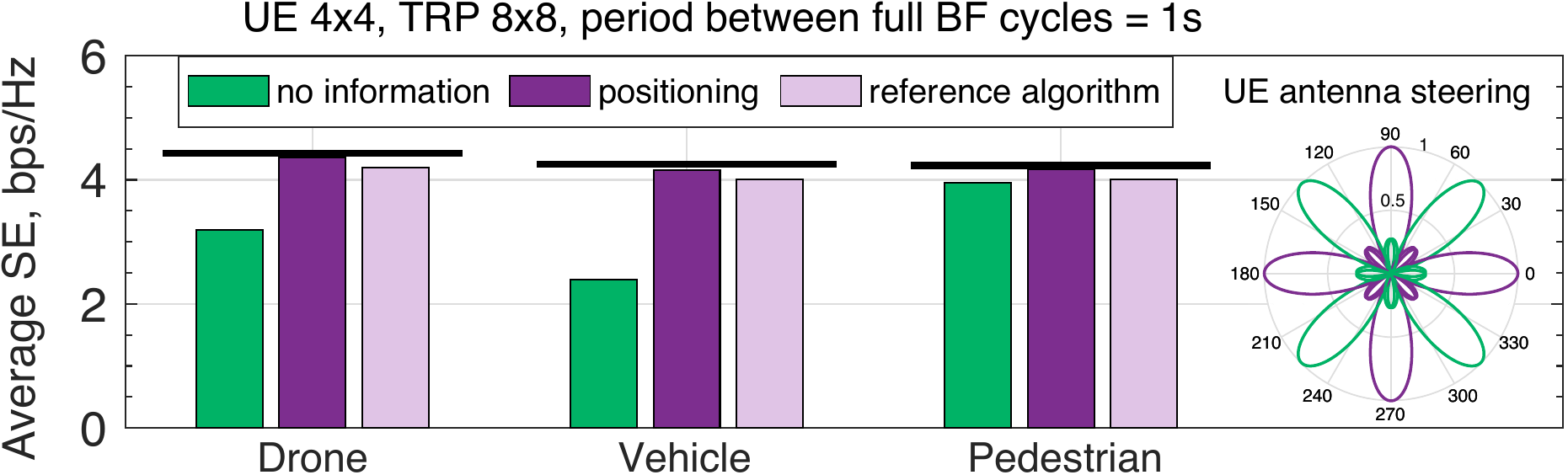}
\caption{Average instantaneous data rate in three scenarios (drone, vehicle, pedestrian) for the beamforming period of 1s} 
\label{fig:results1}
\end{figure}

\begin{figure}[!ht]
\centering
\includegraphics[width=1.0\columnwidth]{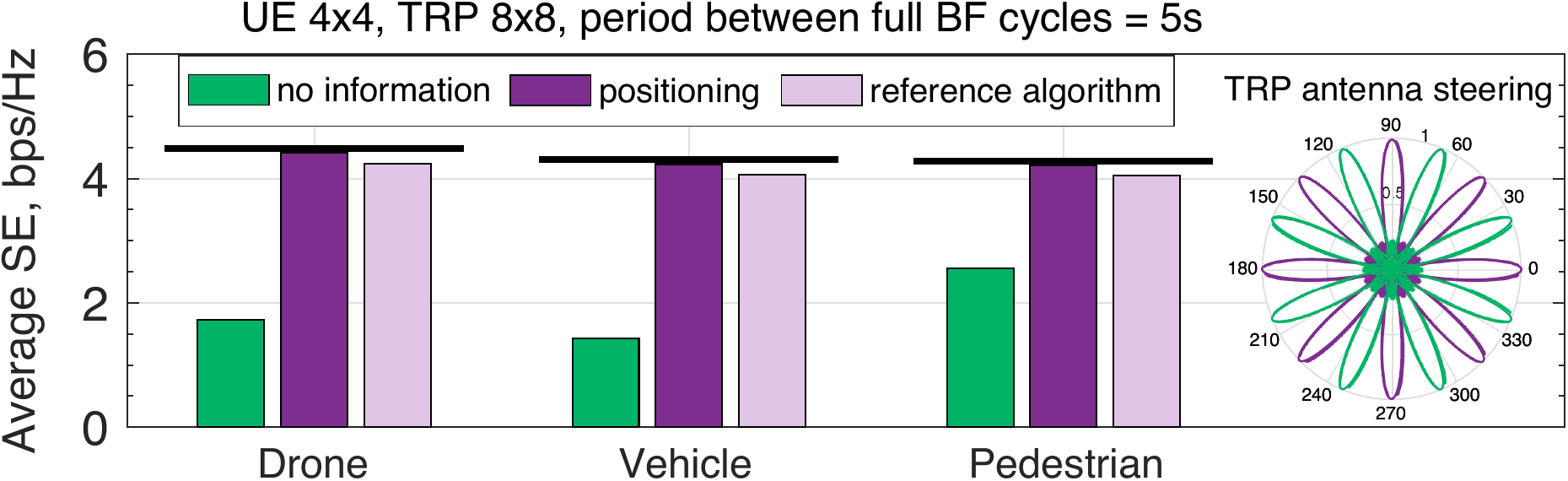}
\caption{Average instantaneous data rate in three scenarios (drone, vehicle, pedestrian) for the beamforming period of 5s} 
\label{fig:results2}
\end{figure}

The utilized shapes of practical beamforming patterns are displayed in Fig.~\ref{fig:results1} and \ref{fig:results2} (right-hand side)~\cite{Tse:2005:FWC:1111206}. The directions of the mmWave beams are then locked, and no beam tracking is conducted further on due to the system implementation complexity considerations. This beam position results in a situation where the SNR and thus the overall system performance may degrade between the following beamforming procedures in case if no positioning information is available. If the TRP is capable of informing the IIoT machines about their current location and orientation (derived as a result of the first phase of our evaluation), both the TRP and the moving device may adjust their beam selection results without the need for additional resource-consuming sweeps.

We assume a certain level of the transmit power, which complies with the current FCC regulations subject to the directivity patterns of the considered mmWave antenna arrays. In our target setup, the IIoT device in question leverages the benefits of multi-connectivity~\cite{7925949} and is thus able to choose the best mmWave link to one of the neighboring TRPs based on the uplink signal strength as proposed in~\cite{giordani2016multi}. Finally, the resulting SNR is translated into the instantaneous data rate via Shannon's formula, which also accounts for the relevant mmWave overheads (including those induced by the beam training) according to our numerology anticipated for 5G mmWave cellular. In summary, Table~\ref{table:parameters} collects the core system-level modeling parameters. 
	
\begin{table}[!ht]
	\centering
	\caption{Main system-level modeling parameters}
	\label{tab:par}
	\begin{tabular}{l|c}
		\hline
		\hline
		\textbf{Parameter}&\textbf{Value}      
		\\ \hline \hline
		Frequency & \SI{28}{GHz}\\
		Channel bandwidth & \SI{100}{GHz}\\
    Sub-carrier spacing & \SI{120}{kHz}\\
		\hline
        Noise figure & \SI{3}{dB}\\
		Default TX power & \SI{200}{mW}\\
		\hline
		Inter-site distance & \SI{25}{m}/\SI{50}{m}\\
        \hline
		Channel model (as per 3GPP TR 38.901) & UMi -- street canyon \\
		\hline
		Number of antenna beams: IIoT device & \SI{8}{}\\
		Number of antenna beams: mmWave TRP & \SI{16}{}\\
		\hline
		Device elevation&\SI{1.2}{m}/\SI{1.5}{m}/\SI{5}{m} \\
		Device speed  & \SI{6}{kmph}/\SI{40}{kmph}/\SI{20}{kmph}\\
		TRP height  & \SI{7}{m}\\
		\hline
    Sweeping subframe length & \SI{0.125}{ms}\\
		Positioning beacon interval & \SI{10}{ms}\\
		\hline   
	\end{tabular}
\label{table:parameters}	
\end{table}

\section{Important Numerical Findings}

\subsection{Position estimation results}

\begin{figure}[!ht]
	\centering
	\includegraphics[width=1.0\columnwidth]{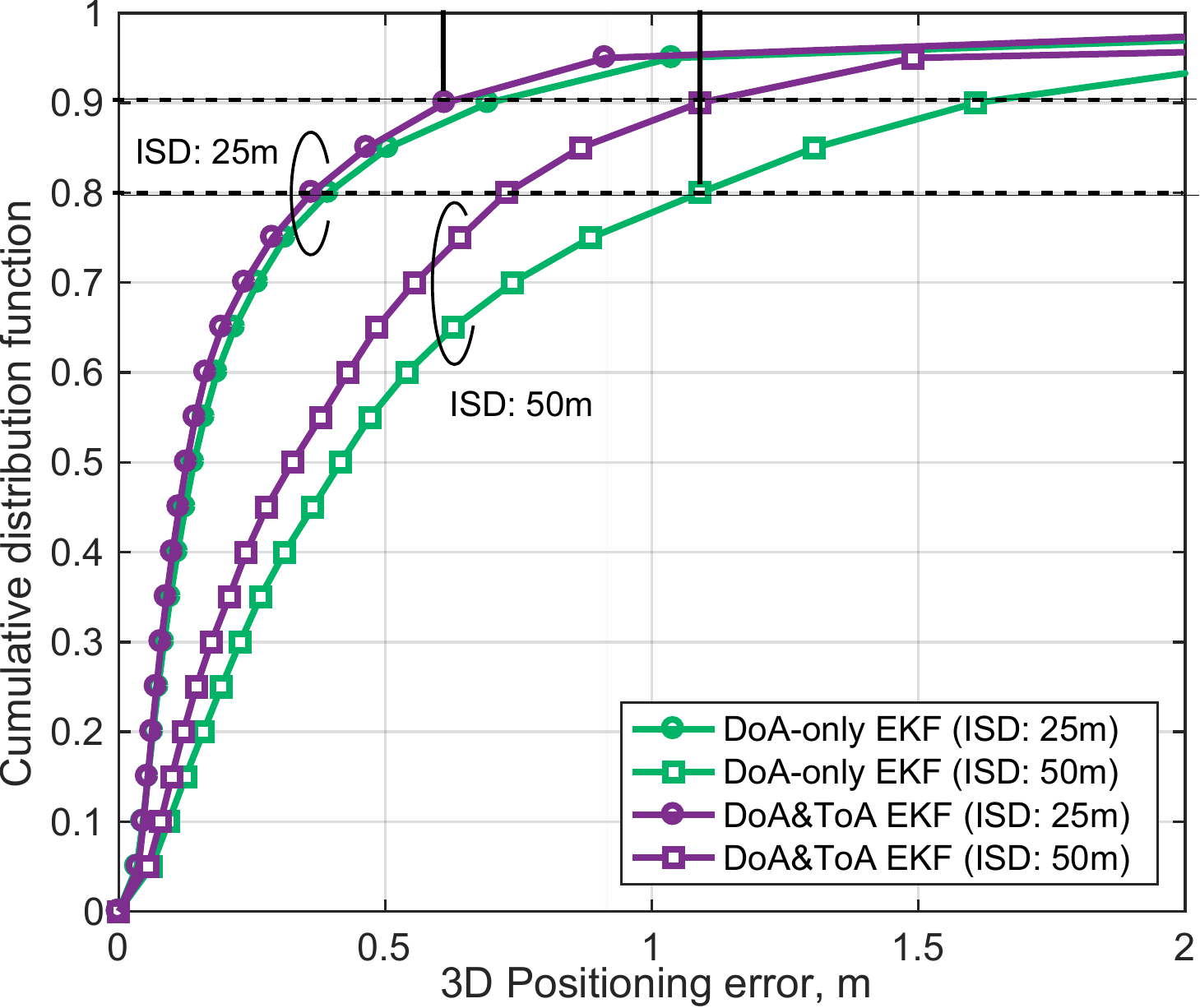}
    \caption{Cumulative Distribution Function (CDF) of 3D positioning errors (in case of, e.g., a drone) for different localization solutions and TRP densities in our 3D positioning setup}
    \label{fig:poscdf}
\end{figure}

We begin our evaluation with a study on the capabilities of our considered 3D positioning algorithms, which is summarized in Fig.~\ref{fig:poscdf}. Here, positioning error distributions are displayed for the two localization approaches discussed above and the ISDs of \SI{25}{m} and \SI{50}{m}. For all the four cases at hand, a positioning accuracy of around \SI{1}{m} may be achieved in more than 80\% of the situations (even in an asynchronous system). As was highlighted previously, this level of localization accuracy is expected to become the minimum requirement for positioning-related functionality in future 5G radio networks. 

Interestingly, our results indicate that the positioning accuracy of as high as \SI{0.5}{m} can be reached in 85\% of the situations when applying the more advanced DoA\&ToA EKF method\footnote{A video demonstration of our 3D drone positioning results is available at http://www.tut.fi/5G/COMMAG18/ [Accessed on 08/2018]} for the ISDs of around \SI{25}{m}. This effect is owing to higher SNR values as compared to the scenarios where the ISDs of about \SI{50}{m} are considered. This, in turn, increases the ToA estimation efficiency and consequently improves the positioning accuracy. 

Generally, higher deployment density of the TRPs may increase the number of handovers of the high-speed target device, hence potentially resulting in a degradation of the positioning performance, especially in an asynchronous network. However, the moving speeds of the considered IIoT machines are relatively low in urban environments, which implies longer connection times for a particular TX-RX pair. In these conditions, the positioning filters are able to capture a sufficient number of localization-related measurements and therefore estimate possible asynchronous effects in such measurements, which enhances the overall performance. 

\subsection{System-level assessment outcomes}

In the second phase, we focus on the following two scenarios: (i) when no additional information is available between the two consecutive beamforming procedures (termed the \textit{baseline} scenario), and (ii) when error-prone positioning information may be utilized at each transmission time interval (termed the \textit{proposed} scenario). The proposed positioning-aided scenario is also compared to another similar approach in~\cite{de2017vision}. In addition, we address a reference \textit{hypothetical} scenario where the exact location of the IIoT device is always known. It serves as an upper bound to benchmark the resulting performance. 

Here, the locations of the TRPs (with the ISD of \SI{50}{m} as an example) are fixed and therefore remain precise. We compare the above-introduced scenarios in terms of the average spectral efficiency for the intervals of \SI{1}{s} and \SI{5}{s} between the periodic beamforming procedures as demonstrated in Fig.~\ref{fig:results1} and \ref{fig:results2}, respectively. For both of these periods, the upper bound (solid horizontal line) slightly decreases with an increase in the average LoS distance between the TX and the RX, which is determined by the actual elevation of the IIoT device. 

The performance of the baseline scheme (green bars) depends heavily on the speed of the target machine, since larger distances traveled without a location update naturally lead to more significant mmWave beam misalignment. As one may expect, the performance improves when the beamforming procedures are invoked more frequently. However, this also increases the overheads dramatically due to long beam training intervals. For the considered periods of \SI{1}{s} and \SI{5}{s}, the overheads are estimated to be on the order of 1-2\%, while for the \SI{10}{s} period they escalate. 

In contrast, our proposed solution (dark violet bars) demonstrates the values comparable to those for the theoretical upper bound, while at the same time keeping the beamforming overheads to a minimum. Another positioning-aided algorithm that we consider~\cite{de2017vision} (mallow bars) indicates a significant improvement with respect to the baseline case. However, it falls behind our proposed solution, and for the case of \SI{1}{s} interval results in a similar performance as that of the baseline pedestrian scenario (Fig.~\ref{fig:results1}). Finally, to understand the evolution of the signal quality between the neighboring TRPs, we additionally consider a dedicated scenario. 

\begin{figure}[!ht]
\centering
\includegraphics[width=1.0\columnwidth]{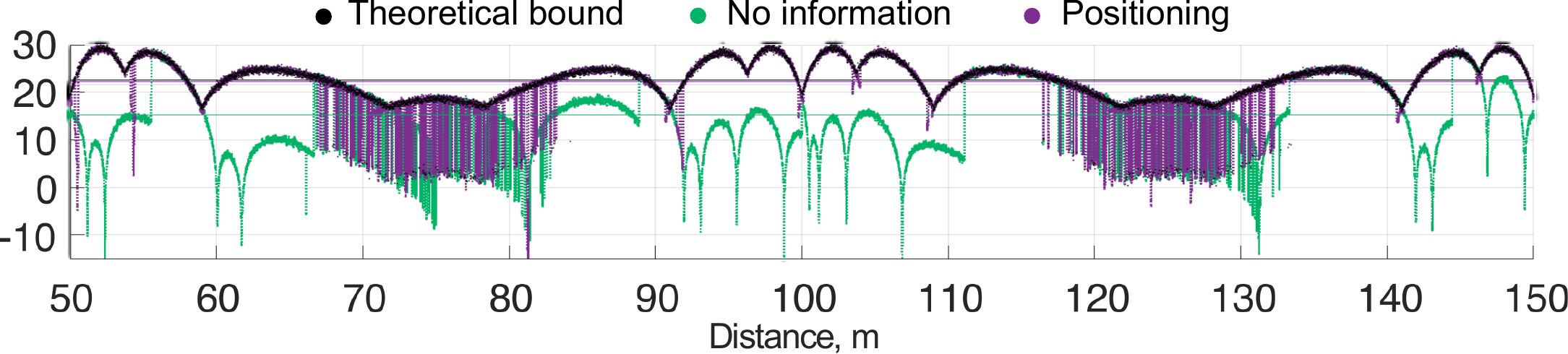}
\caption{System-level evolution of the data rate for, e.g., a drone moving along the line of TRPs, the period is 5s}
\label{fig:results3}
\end{figure}

Accordingly, the IIoT device in question travels on a line of the TRPs separated by the ISD of \SI{50}{m} at the closest distance of \SI{10}{m} to them (see Fig.~\ref{fig:results3} for an illustration of the SNR dynamics in case of a flying drone for the beamforming interval of \SI{5}{s}). Again, the black line corresponds to an upper bound on the SNR (based on the perfect knowledge), while the green curve indicates a considerable quality degradation, especially when the device connects via non-LoS link (3GPP TR 38.901 urban microcell channel model for a street canyon is assumed). Our proposed solution, despite occasional instantaneous drops in quality, rapidly restores the connection and confirms excellent average performance, which remains close to that of the theoretical bound.

\section{Conclusions and Open Research Directions}

In this work, we reviewed the major challenges and potential technology solutions to enable intelligent industry-grade communication of advanced devices (cars, drones, and moving robots) by taking advantage of location awareness and multi-connectivity operation. Our main conclusion is that, in order to achieve this complex target, it is necessary to comprehensively combine the expertise and know-how across the three lines of research: (i) efficient use of mmWave links, (ii) dynamic seamless multi-connectivity, and (iii) positioning-aided computation offloading for demanding industrial applications. More specifically, with our systematic two-phase numerical evaluation, we demonstrated the benefits of DoA and ToA-DoA positioning technologies for improved communication performance of advanced IIoT devices moving at various speeds across an urban landscape. 

The ray-tracing based component of our developed performance assessment methodology allowed us to contrast the operation of alternative location estimation and tracking methods. In our complementary system-level study, we specifically accounted for the advantages of dynamic multi-connectivity at mmWave frequencies (which offer larger available bandwidths) that permitted to achieve seamless communication on the move. Further studies need to be dedicated to improving the indoor positioning accuracy with hybrid cmWave-mmWave signals, as such signals will co-exist in the future IIoT. Another direction is analysis of multi-step evaluation methodologies by taking into account the diverse performance criteria as well as by extending our SNR and spectral efficiency centric results.

\bibliographystyle{ieeetr}
\bibliography{Bib_CommMag}

\section*{Authors' Biographies}

\textbf{Elena Simona Lohan} (elena-simona.lohan@tut.fi) received an M.Sc. degree in Electrical Engineering from Polytechnics University of Bucharest, Romania, in 1997, a D.E.A. degree (French equivalent of master) in Econometrics, at Ecole Polytechnique, Paris, France, in 1998, and a Ph.D. degree in Telecommunications from Tampere University of Technology (TUT), Finland, in 2003. Dr. Lohan is now an Associate Prof. at the Laboratory of Electronics and Communication Engineering (ELT) at TUT. Her current research interests include wireless location techniques, GNSS for aviation, wearables, and privacy-aware positioning solutions.
 
\textbf{Mike Koivisto} (mike.koivisto@tut.fi) received the M.Sc. degree in mathematics from Tampere University of Technology (TUT), Finland, in 2015, where he is currently pursuing the Ph.D. degree. From 2013 to 2016, he was a research assistant with TUT. He is currently a researcher with the Laboratory of Electronics and Communications Engineering, TUT. His research interests include positioning, with an emphasis on network-based positioning and the utilization of location information in future mobile networks.
 
\textbf{Olga Galinina} (olga.galinina@tut.fi) received her B.Sc. and M.Sc. degrees from the Department of Applied Mathematics, St. Petersburg State Polytechnical University of Peter the First, Russia as well as the Ph.D. degree from Tampere University of Technology (TUT). Currently, she is a Finnish Academy Postdoctoral Researcher in the Laboratory of Electronics and Communications Engineering at TUT. Her research interests include applied mathematics and statistics, queueing theory and its applications; wireless networking and energy efficient systems, machine-to-machine and device-to-device communication.

\textbf{Sergey Andreev} (sergey.andreev@tut.fi) received the Specialist and Cand.Sc. degrees from Saint Petersburg State University of Aerospace Instrumentation, Russia, in 2006 and 2009, respectively, and the Ph.D. degree from Tampere University of Technology, Finland, in 2012. He is currently a Senior Research Scientist with the Laboratory of Electronics and Communications Engineering, Tampere University of Technology. He has (co-)authored over 150 published research works on wireless communications, energy efficiency, heterogeneous networking, cooperative communications, and machine-to-machine applications.
 
\textbf{Antti T\"olli} (antti.tolli@oulu.fi) received the Dr.Sc. (Tech.) degree in electrical engineering from the University of Oulu, Oulu, Finland, in 2008. Currently, he also holds an Associate Professor position with the University of Oulu. He has authored more than 160 papers in peer-reviewed international journals and conferences and several patents all in the area of signal processing and wireless communications. His research interests include radio resource management and transceiver design for broadband wireless communications with a special emphasis on distributed interference management in heterogeneous wireless networks.

\textbf{Giuseppe Destino} (giuseppe.destino@oulu.fi) received the Dr.Sc. degree from the University of Oulu, Finland, in 2012, the M.Sc. (EE) degrees simultaneously from the Politecnico di Torino, Italy, and the University of Nice, France, in 2005. He is currently an Academy of Finland Post-Doctoral Researcher and the Project Manager of national and international projects at the Center for Wireless Communications, University of Oulu. His research interests include wireless communications, mm-wave radio access technologies, especially, on algorithms for channel estimation, hybrid beamforming, and positioning.
 
\textbf{M\'ario Costa} (mariocosta@huawei.com) received the M.Sc. degree (Hons.) in communications engineering from the Universidade do Minho, Portugal, in 2008, and the D.Sc. (Tech.) degree in electrical engineering from Aalto University, Finland, in 2013. In 2014, he was a visiting post-doctoral research associate at Princeton University. Since 2014, he has been with Huawei Technologies Oy (Finland) Co., Ltd., as a senior researcher. His research interests include statistical signal processing and wireless communications.
 
\textbf{Kari Lepp\"anen} (kari.leppanen@huawei.com) received the M.Sc. and Ph.D. degrees from Helsinki University of Technology, Finland, in 1992 and 1995, respectively, majoring in space technology and radio engineering. He was with the National Radio Astronomy Observatory, USA, with the Helsinki University of Technology, Finland, with the Joint Institute for VLBI, The Netherlands, and the Nokia Research Center, Finland. He currently leads the 5G Radio Network Technologies Team at Huawei Technologies Oy (Finland) Co., Ltd, Stockholm and Helsinki.
  
\textbf{Yevgeni Koucheryavy} (evgeni.kucheryavy@tut.fi) received the Ph.D. degree from Tampere University of Technology, in 2004. He is currently a Professor with the Laboratory of Electronics and Communications Engineering, Tampere University of Technology, Finland. He is the author of numerous publications in the field of advanced wired and wireless networking and communications. His current research interests include various aspects in heterogeneous wireless communication networks and systems, the Internet of Things and its standardization, and nanocommunications. 

\textbf{Mikko Valkama} (mikko.e.valkama@tut.fi) received his M.Sc. and D.Sc. degrees (both with honors) from Tampere University of Technology, Finland, in 2000 and 2001, respectively. In 2003, he worked as a visiting research fellow at San Diego State University, California. Currently, he is a Full Professor and Head of the Laboratory of Electronics and Communications Engineering at Tampere University of Technology. His research interests include radio communications, radio systems and signal processing, with specific emphasis on 5G and beyond mobile networks.

\end{document}